# A Soft-Switching Single-Stage AC-DC Converter

Farid Naghavi
*Electrical and Computer Engineering Departmenr*
Texas A&M University
College Station, TX
farid@tamu.edu

Hamid Toliyat
*Electrical and Computer Engineering Department*
Texas A&M University
College Station, TX
toliyat@tamu.edu

*Abstract*—Partial resonance high frequency AC link converters are recognized to provide soft-switching, single-stage power conversion. They are highly reliable due to elimination of DC capacitors. Previous work has focused on proof of concept topologies and design requirements. In this work, a current regulated AC-DC converter from this family of converters is presented. The control design and grid integration requirements are discussed. An active damping method using grid current feedback is implemented to mitigate the filter resonance oscillations. The current control is implemented in the synchronous frame using voltage oriented control which enable power factor and power flow control. The result of converter operation and its control loops' performances are presented.

## I. Introduction (Heading 1)

Active rectifiers or AC/DC converters are widely used in battery chargers, motor drives, renewable energy systems, microgrids, active filters, and solid state transformer [1]–[10]. These converter have certain advantages compared to passive rectifiers such as power factor control, lower current THD, power flow control and output voltage regulation [1], [6], [11]-[12].

Typical topologies used for active AC/DC rectifiers are buck-type, boost-type and the voltage source rectifier (VSR) [6], [11]-[13] and current source rectifiers (CSR) [7], [13], [14]. In a VSR, the output voltage level is limited by the line-line voltage of the input meaning that for high voltage ratios, a DC-DC converter should be used to change the voltage level [11], [12], [15]- [16]. If isolation is required, an isolated DC-DC converter or a line transformer can be used. [4], [5], [11], [12], [20], which result in a multi-stage solution which becomes complex in design and control. Each stage has to be highly efficient for the system to achieve a high efficiency as the series connection of stages can deteriorate the efficiency.

Partial resonance high frequency AC link converter perform the power conversion in a single stage. They have the ability to buck or boos the voltage. In addition by taking advantage of partial resonance of an AC link soft-switching is achieved. Soft-switching is achieved for all the operating points of the converter without any limitations. Furthermore, due to the elimination of DC electrolytic capacitors they are highly reliable. Isolation is provided by means of a high frequency transformer (Fig. 1). Previous work [21]-[26] have mostly focused on proof of concept topologies and design guidelines for this family of converters however, this paper focuses on the current regulation implementation and grid-connection process for an AC/DC converter.

Common control methods for AC/DC converters are Voltage-oriented control (VOC) and direct power control (DPC) [1], [17]-[18]. In VOC, the active and reactive power are controlled indirectly by controlling the current. When the current and voltage vectors are in phase, unity power factor is achieved [1], [18]. The internal current controller can be implemented using PR controller in stationary frame or using a PI controller in the synchronous frame [4], [19]. In DPC, active and reactive power are calculated using the voltage and current measurements and the converter is controlled using a hysteresis or bang-bang controller and a switching table to directly control active and reactive power [1], [17]-[18].

Fig. 1, shows an AC Link AC-DC converter. The AC Link converter employs a CL filter on the input to mitigate current harmonics generated by switching of the converter and also to facilitate soft switching. Due to the low damping of the CL filter resonance oscillations and instabilities can occur due to low. An active damping control or virtual impedance [14], [27]-[29] has been implemented to solve this problem. The active damping loop, increases the damping of the filter by modifying the converter reference current without physically adding a resistor to the filter.

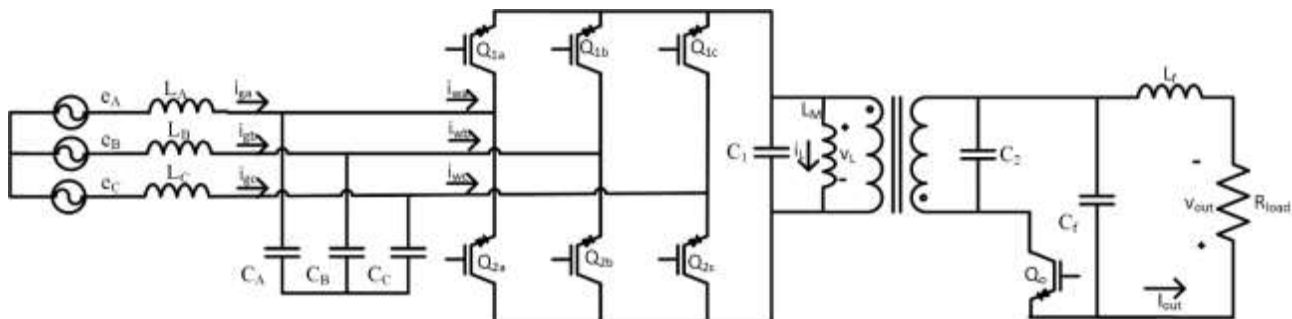
Fig. 1 AC-DC AC Link converter

The rest of this paper is organized as follows: The converter's principles of operation is discussed in section II, the current control and active damping loop are analyzed in section III and IV. The results are report in section V and section VI conclude the paper.

## II. AC Link Rectifier Operation Principles

The partial resonance AC link AC-DC converter utilizes reverse-blocking switches similar to a current-source converter. The AC link consists of the magnetizing inductance of the transformer ($L_M$) and the link capacitors ($C_1$ and $C_2$). The magnetizing inductance charges through the input phases and then disconnected from the input. A resonance circuit is formed by the magnetizing inductance and the link capacitors but the resonance is only allowed to happen partially and the stored energy in the magnetizing inductance is discharged into the output at appropriate times.

### A. Switch Controller

The converter's switches are controlled through a sequencer that generates the gate signals based on the link current, link voltage, input reference currents and output reference current. The switch controller in essence maintains the input and output energy balance for every link cycle. The input references and the output reference are generated by the higher level current controller (Fig. 4). The switch controller commands when to start and end each mode of operation based on the current and voltage measurements in order to satisfy the conditions for soft switching. The operation modes of the converter are detailed in the next section. The switch controller can be considered as the equivalent of the modulator in a typical PWM converter.

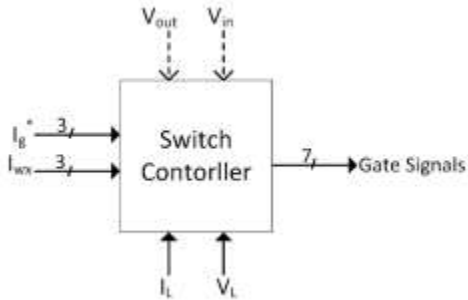

Fig. 2

### B. Operation Modes

The converter has 6 modes of operation which consist of 3 energy transfer modes and 3 partial resonance modes. The high level current controller determines the current references for the input and output currents. The switch controller decides when to start and end each operating mode based on the measurements to facilitate soft-switching.

*Mode 1:* The link voltage has peaked prior to mode 1 and will be decreasing in this mode. The link peak voltage is higher than the maximum input line-line voltages. Therefore, when the link voltage is at its peak all the input switches are reverse biased and cannot conduct.

The link is charged in modes 1 and 3 by connecting to the input. The current reference for each phase is determined by the higher level current controller. One switch from each phase is selected for modes 1 and 3. For instance, if the current references are $I_a$=1A, $I_b$=-0.5A and $I_c$=-0.5A, switches $Q_{1a}$, $Q_{2b}$ and $Q_{2c}$ should conduct during modes 1 and 3 and the corresponding phase pairs are AB and AC meaning phase A will conduct in both modes 1 and 3 and phase B and C conduct only in one mode.

All the switches are reverse biased when the link voltage is at its peak so if the gate signals for these three switches are generated, the switches stay off until they are forward biased. As the link voltage decreases, it reaches one of the phase pairs. At that moment two of the switches start to conduct (e.g. A and C). The switch controller monitors the currents and as soon as the average current of a phase (e.g. C) is equal to its reference the switch for that phase is tuned off.

At the moment that the switches in mode 1 start conducting, the link voltage is equal to the selected phase pair line-line voltage ($V_{ac}$ or -$V_{ca}$) meaning that the voltage drop across the switches is zero hence the switches turn-on with zero voltage switching (ZVS). On the other hand, when one of the switches turn off at the end of mode 1, due to the filer capacitors at the input ($C_A$-$C_C$) and the link capacitor ($C_1$), the voltage across the switch is almost zero. As a result the turn off is also ZVS. This soft-switching mechanism in maintained for all the switches at all operating points of the converter.

*Mode 2:* After the first switch is turned off at the end of mode 1, the link enters a partial resonance mode and the link voltage continues to decrease. None of the switches conduct in this mode since they are reverse biased.

*Mode3:* As the link voltage decreases, it will be equal to the other phase pair (e.g. AB) and at this moment the switches for the other phase pair (e.g. $Q_{1a}$ and $Q_{2b}$) become forward biased and start to conduct. Mode 3 continues until the references for the other two phases are met. Since the sum of the three phase currents is zero, the other two currents references are met simultaneously. All the input switches are turned off at the end of mode 3.

*Mode 4:* Since all the switches are off in mode 4, the converter enters another partial resonance mode. The link voltage continues to decrease in this mode and becomes negative. The link current also reaches its peak when the link voltage is zero.

*Mode 5:* The link is allowed to resonance until the output switch is forward biased (-$V_L$>$V_{out}$). At this moment, the output switch starts to conduct and discharge the stored energy into the output.

The switch controller monitors the link energy in this mode to make sure there is a small amount of residual energy left in the link so it can reach the predetermined peak voltage again. The link has to reach this peak to satisfy the condition for soft-switching. As discussed in previous modes, the descending order of the link charging through the input phases is essential to the soft-switching mechanism.

If the reference for this mode (either voltage or current) is not met the output controller adjusts the input reference ($I_g^*$) so that the output voltage or current is regulated (Fig. 5).

*Mode 6:* This mode is another partial resonance mode that continues until the converter enters mode 1 again.

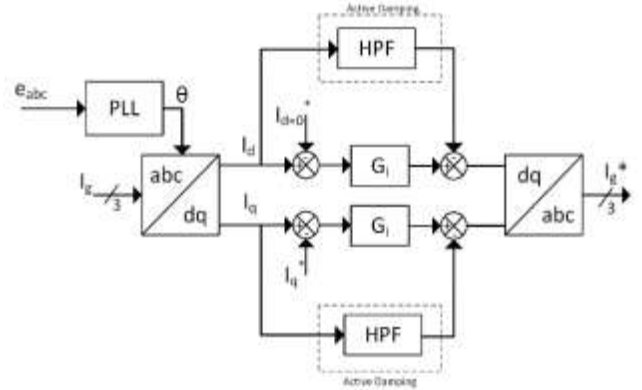

Fig. 4 Current control block diagram

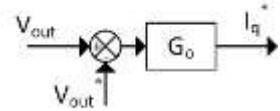

Fig. 5 Output controller block diagram

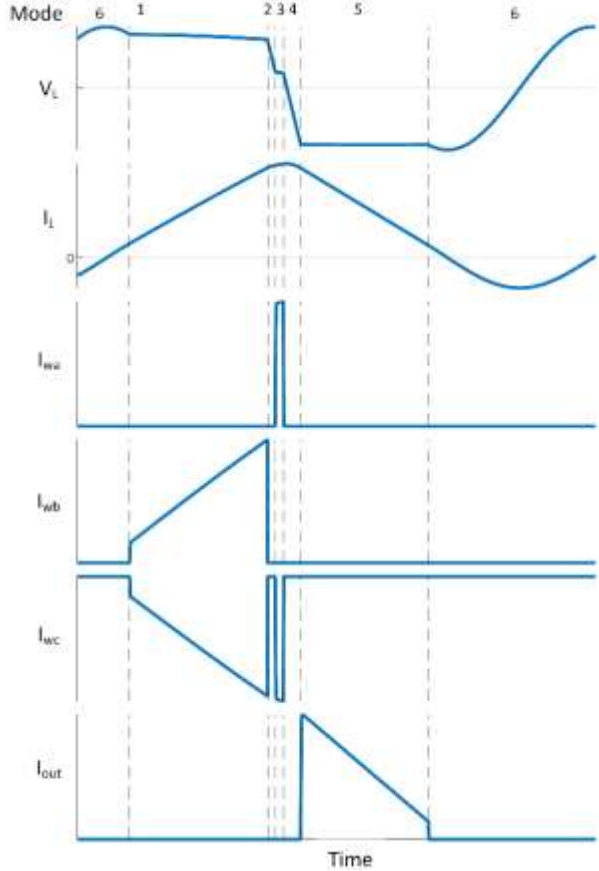

Fig. 3 A typical link cycle of the converter

## III. CONTROL STATEGY

### A. Current Control

Fig. 4, shows the overall control block diagram using VOC. A phase-locked loop (PLL) extracts the grid voltage angle. The angle from the PLL is used as the reference angle for transformation of the abc grid currents ($I_{gabc}$) to the synchronous frame (dq). Based on the convention used here, the current in d-axis ($I_d$) controls the reactive power and the q-axis current ($I_q$) controls the active power.

The reference for d-axis current is set to zero ($I_d^*=0$) to achieve unity power factor. The reference current in the q-axis is set by the output controller as shown in Fig. 5. The parameter that is controlled at the output depend on the load it can be either the voltage or the current or both.

A PI compensator is used in dq axes to regulate the current. The dq current are then transformed back to the abc frame and then fed to the switch controller. An alternative is to control the current in the stationary frame using p+resonance controllers. A PI controller in the abc frame can suffer from steady state error.

### B. Active Damping

The input CL filter at the input of the converter filters the switching components and also provides the conditions for soft-switching. Since the series resistance of the filter inductor and capacitor are small, the damping in the CL filter is low. As a result, resonance oscillations can happen that distort the current and increases the current THD. In addition, based on the bandwidth of the current loop, it can also cause instability in the current loop.

In order to dampen the oscillation, an active damping or virtual impedance method is used. The reference grid current is manipulated in such a way to emulate an actual physical resistor that is placed in series or parallel with the CL filter. In contrast to adding a physical resistor, active damping does not decrease the efficiency of the converter. There are a few different techniques reported to implement the active damping such as using capacitor voltage feedback or inductor current feedback [27]–[32]. It can be implemented in the synchronous frame or the stationary frame.

In this work the active damping is implemented using inductor feedback in the stationary frame. The active damping block is shown in Fig. 4 with dashed lines. A high-pass filter (HPF) blocks the dc components of the inductor current in dq-axis and only the higher order components that involve the resonance frequency are passed. The non-dc components of the inductor current are subtracted from the current controller ($G_i$) output before being transformed to the abc frame. Consequently, the reference grid current is modified to cancel out the current harmonics related to the CL filter resonance frequency.

## IV. PARAMETER DESIGN

### A. Active Damping Loop

The CL filter is a 2$^{nd}$ order system with low damping. The gird to converter current transfer function without active damping is

$$G_p(s) = \frac{I_g(s)}{I_w(s)} = \frac{1}{LCs^2 + r_sCs + 1} \quad (1)$$

Fig. 6, shows the transfer function without any active damping. It can be seen that the transfer function at the resonance frequency becomes unstable. The active damping modifies this transfer function to alter the frequency response. The active damping block is an HPF with the form:

$$HPF(s) = \frac{ks}{1 + \frac{s}{\omega_c}} \quad (2)$$

k is the damping factor and is selected according to (3) based on the filter inductor and capacitor values and the desired damping coefficient $\xi$ [32]. The dq components of the current fundamental frequency (60Hz) are DC quantities in the synchronous frame and are block by the HPF. The filter cut-off frequency ($\omega_c$) should be at least twice the resonance frequency to effectively extract high frequency components of the current [32].

$$k = 2\xi\sqrt{LC} \quad (3)$$

The input current transfer function with the active damping loop can be calculated using (4).

$$G_{ig}(s) = \frac{I_g(s)}{I_w(s)} = \frac{G_p(s)}{1 + G_p(s) * HPF(s)} \quad (4)$$

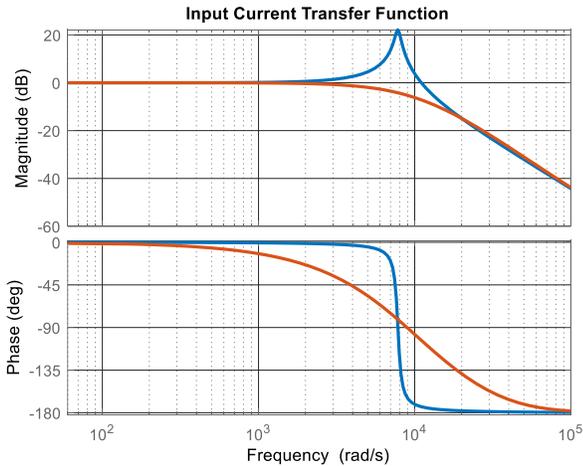

Fig. 6 Bode plots of grid current transfer function blue: without active damping red: with active damping

### B. Current Controller

Due to the fact that the dq currents in the synchronous frame are DC quantities, a PI regulator can regulate the current with zero steady state error. The current controller bandwidth is selected to be 1/5$^{th}$ of the inner loop crossover frequency to prevent any interference. Fig. 7, shows the closed-loop response of the grid current.

$$G_i(s) = \frac{K_p + K_i s}{s} \quad (5)$$

Fig. 7, shows the Bode plot of the loop gain (6) with parameters listed in Table I.

$$G_L(s) = \frac{I_{dq}^*}{I_{dq}} = G_{ig}(s)G_i(s) \quad (6)$$

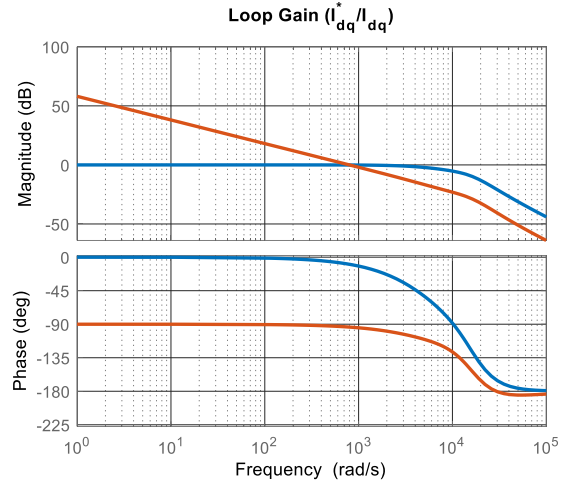

Fig. 7 Bode plots of loop gain (red) and inner active damping loop (blue)

## V. RESULTS

A 500W converter with parameters in Table I, is analyzed and simulated.

TABLE I
CONVERTER PARAMETERS

| | |
|---|---|
| Input Voltage (V) | 80 |
| Input Current (A) | 3.6 |
| Output Voltage (V) | 50-150 |
| Magnetizing Inductance (µH) | 425 |
| Link capacitors(nF) | 100 |
| Input Filter Inductance, L$_f$ (mH) | 1.6 |
| Input Filter Capacitance (µF) | 40 |
| Damping Factor, k | 0.0003 |
| HPF cut-off frequency (Hz) | 3000 |
| K$_i$ | 800 |
| K$_p$ | 0.1 |

Fig. 8, shows the grid current with the active damping loop disabled. The grid current has a THD of 39.2% due to the CL filter resonance. From grid current FFT, as shown in Fig. 9, it can be seen that the resonance harmonics are the main cause for the high THD.

Fig. 10 and 11 show the grid current and its FFT when the active damping is enabled. The reference is set to 2A (I$_q$=2A and I$_d$=0A). The initial oscillation is due to the transient delay associated with the HPF in the active damping loop. The current THD is reduced to 1.88%.

Fig. 12, shows the grid voltage and current for phase A. As can be seen the voltage and current are in phase, therefore, the controller is capable to regulate the d-axis current to zero and maintain unity power factor.

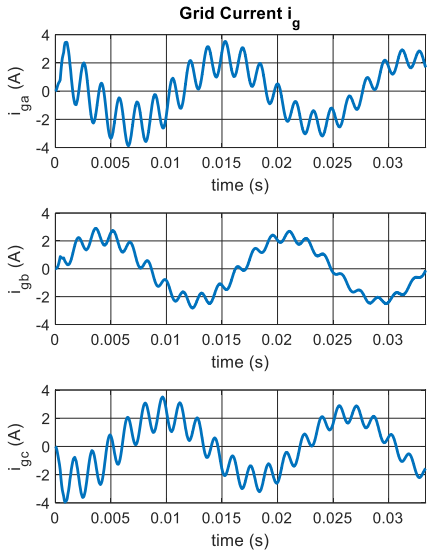

Fig. 8 Grid current without active damping

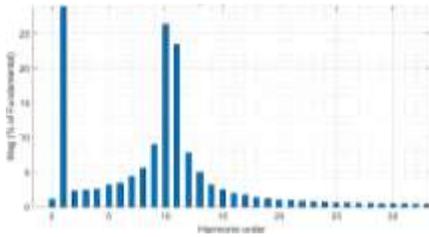

Fig. 9 FFT of grid current without active damping

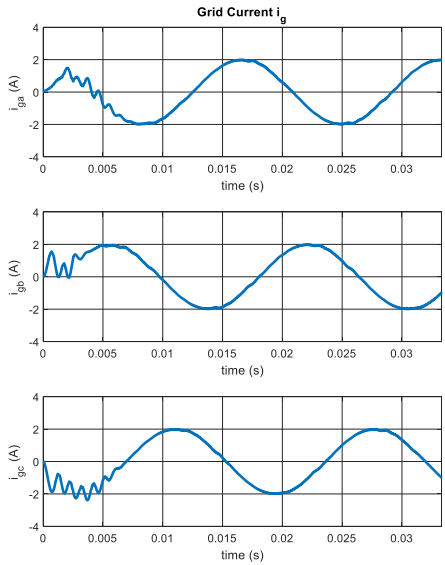

Fig. 10 Grid current startup with $I_q=2$ and $I_d=0$

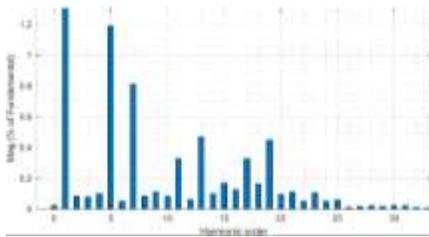

Fig. 11 Grid current FFT with active damping enabled

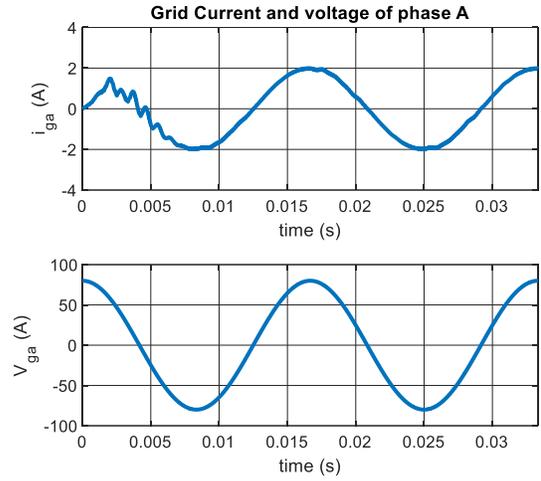

Fig. 12 Phase A Voltage and Current showing unity power factor operation

Fig. 13 shows the current response to a reference change from 2A to 4A. Fig. 14, shows the operation of the outer loop that regulated the output voltage. It shows the grid currents and the output voltage at startup of the converter.

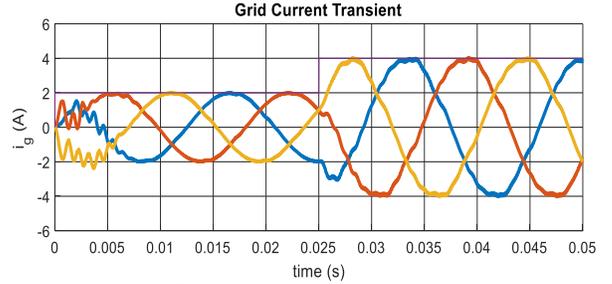

Fig. 13 Grid current transient

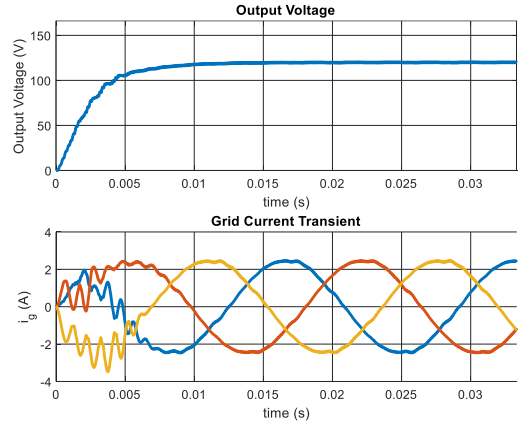

Fig. 14 Output voltage response

VI. CONCLUSION

A partial resonance AC link AC-DC converter with soft-switching is analyzed in this paper. An advantage of this converter is the soft-switching of all the switching at all operating conditions. The control requirement and design procedure was discussed. Because of the CL filter on the input side of the converter, an active damping technique is used to mitigate the oscillations. The current control is implemented in the synchronous frame using VOC. The results of the regulated converter are shown and the control loop operation verified.